CAMBRIDGE
UNIVERSITY PRESS

# Data-driven surrogate modeling and benchmarking for process equipment

Gabriel F. N. Gonçalves[1] , Assen Batchvarov[1], Yuyi Liu[1], Yuxin Liu[1], Lachlan R. Mason[2], Indranil Pan[2,3] and Omar K. Matar[1,*]

[1]Department of Chemical Engineering, Imperial College London, London, United Kingdom
[2]Data Centric Engineering Program, The Alan Turing Institute, London, United Kingdom
[3]Centre for Environmental Policy, Imperial College London, London, United Kingdom
*Corresponding author. E-mail: o.matar@imperial.ac.uk

## Abstract

In chemical process engineering, surrogate models of complex systems are often necessary for tasks of domain exploration, sensitivity analysis of the design parameters, and optimization. A suite of computational fluid dynamics (CFD) simulations geared toward chemical process equipment modeling has been developed and validated with experimental results from the literature. Various regression-based active learning strategies are explored with these CFD simulators *in-the-loop* under the constraints of a limited function evaluation budget. Specifically, five different sampling strategies and five regression techniques are compared, considering a set of four test cases of industrial significance and varying complexity. Gaussian process regression was observed to have a consistently good performance for these applications. The present quantitative study outlines the pros and cons of the different available techniques and highlights the best practices for their adoption. The test cases and tools are available with an open-source license to ensure reproducibility and engage the wider research community in contributing to both the CFD models and developing and benchmarking new improved algorithms tailored to this field.

### Impact Statement

The recommendations provided here can be used for engineers interested in building computationally inexpensive surrogate models for fluid systems for design or optimization purposes. The test cases can be used by researchers to test and benchmark new algorithms for active learning for this class of problems. An open-source library with tools and scripts has been provided in order to support derived work.

## 1. Introduction

Process models are used in the design and operation stages of a process plant for a wide array of tasks including design optimization, sensitivity analysis and uncertainty quantification among others (McBride and Sundmacher, 2019). Most common process models use regression, where a variable of interest (e.g., a drag, or heat transfer coefficient) can be mapped as some function of operating conditions, fluid properties, and/or geometric parameters. Usually, experimental tests are performed in down-scaled versions of the system of interest, and the measurements are used to develop regression functions on an ad hoc basis. Computational fluid dynamics (CFD) simulations are an attractive tool to complement such a methodology as once a model has been set up and validated, parametric analysis can be easily performed manually or automatically. Additionally, numerical computations allow for the exploration of parameters that could not be easily replicated in laboratory conditions due to infrastructure or measurement limitations.

Despite the aforementioned advantages, realistic CFD simulations suffer from major limitations in terms of computational cost. Simulating moderately sized industrial systems may take hours to days on high-performance computing clusters, which means that in practice the total number of simulations in a given study will rarely be greater than a few dozen. The regression algorithm should therefore have the highest possible sample efficiency, that is, be able to provide good estimates of interpolated values with the lowest number of simulations possible. Traditional techniques used in the design of computational experiments include Latin hypercubes, Hammersley sequence sampling, orthogonal arrays and uniform designs (Simpson et al., 2001). These methods aim to distribute the input features such that the space of parameters is filled as "evenly" as possible. In the context of machine learning, different strategies have been used for dealing with small sample regression problems where obtaining samples is costly. They include transfer learning, semi-supervised learning, and active learning (Wu et al., 2019) among others. This study will focus on the last strategy, since it requires minimal interference on the surrogate development algorithm and is also agnostic of the underlying CFD model. Note, however, that these techniques are not necessarily exclusive, and combinations of methods could also be explored for further improvements.

Active learning consists of a mathematical criterion, a sampling strategy, to select the most "informational" parameters from a pool of possibilities. In the present case, this consists of selecting the next expensive simulation to be performed and is decided algorithmically, possibly making use of the underlying structure of the solution estimated with previous results. This avoids an extremely expensive grid search of all parametric permutations, a procedure that is infeasible in moderately complex systems. Such grid search techniques suffer from the "curse of dimensionality," where the number of possible combinations of simulations increases exponentially with respect to the increase in dimensionality, making such techniques computationally intractable. Active learning techniques for regression problems have been the subject of relatively few studies, when compared to classification tasks. Ray-Chaudhuri and Hamey (2002) utilized the "query-by-committee" scheme for active learning with a neural network. Several copies of the model were trained on random subsets of the data and the discrepancy of the predictions was used as a criterion for selecting samples. Cai et al. (2013) proposed an active learning algorithm which selects the examples expected to affect the predictions the most. The expected change is estimated based on the comparison between models, which were trained in subsets of the data. O'Neil (2015) compared active learning techniques in the context of linear regression models. From all the schemes investigated, only the diversity-based approach was found to significantly outperform the baseline results. Wu et al. (2019) analyzed the performance of six sample selection algorithms for regression, of which two were proposed by the author, on datasets from various domains. The five active learning techniques were found to outperform random selection consistently. The strategy based on greedy sampling of both inputs and output was verified to be particularly effective.

However, there does not seem to be a generic active learning strategy which can outperform all others for every class of problems. On the other hand, empirical and systematic comparison of algorithms in different domains of knowledge may be useful for a practitioner interested in applying such techniques to a specific class of problems. The present idea is to build a set of industrially relevant problems which are generic enough to cover similar types of CFD problems and benchmark different active learning algorithms to assess their efficacy and provide directions for best practices to users in similar domains. In the present work, a methodology for generation of surrogate models based on CFD simulations is explored and evaluated with a set of test cases from the literature. The library developed for benchmarking is available on GitHub[1] under an MIT Licence, which allows users to freely "use, copy, modify, merge, publish, distribute, sublicense, and/or sell copies of the software." This includes functions for selecting between different strategies, preparing and extracting results from CFD cases and calculating errors, as well as case setups and plotting scripts.

---

[1] https://github.com/ImperialCollegeLondon/al_cfd_benchmark.

The remainder of this article is organized as follows: in Section 2, the surrogate model development algorithm will be presented, along with the specific regression and sampling techniques which were evaluated. Following that, Section 3 will introduce the test cases used in the present work and Section 4.1 will detail the methodology used for comparison between methods. Finally, Section 4.2 will present the results of the analysis, along with general conclusions and recommendations for the engineering applicability of the methodology.

## 2. Surrogate Modeling Framework

The objective of the surrogate model approach is to develop computationally inexpensive statistical model which, following a systematic calibration can reproduce key predictions of a complete CFD simulation at a fraction of the computational cost. Mathematically, the aim is to find a regressor $f^{\star}(\mathbf{x})$ that approximates a function $f(\mathbf{x})$ for the domain $\mathbf{x} \in [\mathbf{x}_{\min}, \mathbf{x}_{\max}]$, based on a limited number $N_s$ of function evaluations $\{f(\mathbf{x}_i)\}, i \in [1, N_s]$. Since the toolbox is designed generically to evaluate the sampling criterion and choose from a discrete *pool* of parameters, the domain is divided into a fine grid of possible simulation conditions. An initial set of working conditions $\{\mathbf{x}_i\}, i \in [1, N_0]$ is chosen with a fixed initialization strategy, and the respective simulations are performed in order to obtain a set of initial responses. A regression model is built based on these results. Following that, a sampling algorithm selects a new operating condition (e.g., from the available *pool*) to be evaluated. This is repeated until $N_s$ simulations have been performed. This workflow is shown in Figure 1. Alternatively the stopping criterion could be defined based on some measure of estimated uncertainty. For benchmarking purposes, however, the procedure presented here seems more appropriate, allowing for a direct comparison between different methods given a fixed number of simulations.

### 2.1. Regression techniques

Five regression techniques of varying complexity were chosen for evaluation. The key concepts will be introduced here; for details on each model we refer the reader to Hastie et al. (2009). The implementations were provided in the *scikit-learn* package (Pedregosa et al., 2011). Most of these techniques depend on *hyperparameters*—parameters the values of which are set before the training process begins. These need to be adjusted through empirical rules and systematic testing. In order to keep the procedure of this work simple and relevant to engineering applications, we forfeited fine-tuning of default parameters whenever possible, restricting the analysis to the defaults provided by the software. These settings have been shown to work well across a wide range of problem domains and have hence been recommended as defaults in the popular software implementations. As such, they would also be the default choice to a user applying such active learning techniques to a new CFD model with minimal prior insight regarding the choice of hyperparameters. Therefore, the results of each method should not be seen as best case scenarios, but instead as a typical results.

#### 2.1.1. Linear regression

The most basic kind of regression is a linear model, given by:

$$f^{\star}(\mathbf{x}) = \mathbf{a} \cdot \mathbf{x} + b, \tag{1}$$

where $\mathbf{a}$ and $b$ are coefficients to be adjusted based on observations.

Linear regression can often describe the behavior of functions over small ranges, but eventually becomes too limited to describe data from real systems. It is included here as a baseline for comparison to more sophisticated models.

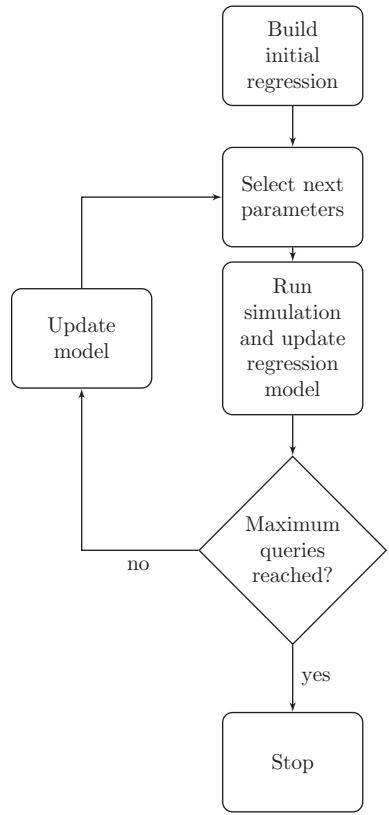

**Figure 1.** *Workflow utilized for building the surrogate models for computational fluid dynamics (CFD) simulations through the active learning methodology.*

### 2.1.2. Gaussian process regression

Gaussian process (GP) regression is a nonparametric Bayesian technique, which provides the regression as a distribution over functions. As such, predictions are also characterized by a mean value and a standard deviation. The mean values of the predictions are given by:

$$f^{\star}(\mathbf{x}) = \sum_{i=0}^{N_t} w_i k(\mathbf{x}_i, \mathbf{x}), \tag{2}$$

where $\mathbf{x}_i$ are the parameters used for training and $N_t$ is the total number of observations. The weights $w_i$ can be calculated with the knowledge of the kernel function $k(\mathbf{x}_i \mathbf{x}_j)$ and the responses $f(\mathbf{x}_i)$. A kernel function $k(\mathbf{x}_i \mathbf{x}_j)$ must be chosen to generate the covariance matrix, which defines the smoothness of the regression function. Here, three alternatives will be compared: the radial-basis function (RBF), Matérn, and cubic kernels (Rasmussen and Williams, 2005). The RBF kernel is a popular choice for GP models, given by:

$$k(\mathbf{x}_i, \mathbf{x}_j) = \exp\left(-\frac{1}{2}d(\mathbf{x}_i/l, \mathbf{x}_j/l)^2\right), \tag{3}$$

where $d$ is the Euclidean distance between the vectors and $l$ is a length-scale hyperparameter, which should be strictly positive. The Matérn kernel is a generalization of the RBF kernel. It is defined by an additional parameter $v_M$. In the particular case of $v_M = 5/2$, the obtained functions are twice differentiable, and the kernel is given by:

$$k(\mathbf{x}_i, \mathbf{x}_j) = \left(1 + \sqrt{5}d(\mathbf{x}_i/l, \mathbf{x}_j/l) + \frac{5}{3}d(\mathbf{x}_i/l, \mathbf{x}_j/l)^2\right)\exp\left(-\sqrt{5}d(\mathbf{x}_i/l, \mathbf{x}_j/l)\right). \tag{4}$$

In the present models, the initial length scale $l$ was set to 0.1. The cubic kernel is given by:

$$k\left(\mathbf{x}_i, \mathbf{x}_j\right) = \left(\sigma_0^2 + \mathbf{x}_i \cdot \mathbf{x}_j\right)^3, \tag{5}$$

where $\sigma_0$ is a hyperparameter. The parameter estimation was performed using an optimization algorithm (L-BFGS-B, a variation of the Limited-memory Broyden–Fletcher–Goldfarb–Shanno algorithm; Zhu et al., 1997).

### 2.1.3. Random forest regression

In a random forest regression, the response is obtained from a combination of the predictions of several decision trees, which are trained on random subsets of the complete dataset. The technique has become very popular for nonlinear regression, scaling well to many input parameters and large amounts of data. It also allows for an inspection of the importance of individual features in the results, which can be useful for interpretability. The number of trees was set to 10.

### 2.1.4. Support vector regression

Support vector regression relies on projecting the input data onto a reproducing kernel Hilbert space, where a linear regression is then performed. Here, the frequently adopted RBF kernel (Equation 3) was used.

### 2.1.5. Multilayer perceptron

A multilayer perceptron is a type of neural network composed of a sequence of layers of neurons. A series of linear operations and nonlinear activation functions is applied to the input vector as it passes through the network, until a single output vector is obtained. Since data are scarce in the cases considered here, a single hidden layer with 10 neurons was used. The stochastic optimization algorithm *Adam* (Kingma and Ba, 2014) was used for weight adjustment.

### 2.1.6. Summary

A summary of the total number of adjustable parameters for each technique, with the configurations presented before, is shown in Table 1. Here, $N_f$ is the number of features in the input vector. In a situation of scarce data, as examined here, the computational cost of fitting the parameters is negligible in comparison to the data collection. In scenarios of abundant data, on the other hand, the scalability of each method in respect to $N_t$ may be a matter of concern.

## 2.2. Sampling strategies

Five sampling strategies were compared: random sampling, three variations of greedy sampling presented by Wu et al. (2019) and a strategy based on the estimation of prediction variance, specific to GP models. The active learning strategies were implemented using the *modAL* Python package (Danka and Horvath, 2018).

**Table 1.** Number of fitting parameters used in each regression technique.

| Method | Number of parameters |
|---|---|
| Linear | $N_f$ coefficients, 1 intercept |
| GP | $N_t$ coefficients, 1 hyperparameter |
| Random forest | 10 decision trees, variable number of leaves |
| Support vector | $N_t$ coefficients, 1 intercept |
| Multilayer perceptron | $10(N_f + 1)$ weights, 11 intercepts |

### 2.2.1. Random sampling

Random sampling is commonly used as a base reference for comparing against other methodologies. Since it does not require any kind of modeling, it has negligible computational overhead, and is readily parallelizable. However, the fact that it neglects the underlying structure of the response space means that for complex functions it might require a very large number of function evaluations to converge. Samples are randomly chosen from the pool:

$$\{\mathbf{x}_i\} \xrightarrow{R} \mathbf{x}_s. \tag{6}$$

### 2.2.2. Greedy sampling on the inputs

Greedy sampling on the inputs seeks to sample the locations which maximize the exploration of the input space, ignoring outputs. It is not dependent on a regression model for estimation of outputs, which means that the computational overhead is low. The distance $d$ in feature space of an unlabeled instance $\mathbf{x}_i$ to the already labeled positions $\mathbf{x}_j$ is given by:

$$d_{\mathbf{x}}(\mathbf{x}_i) = \min_j \|\mathbf{x}_i - \mathbf{x}_j\|. \tag{7}$$

For greedy sampling on inputs, each sample is selected according to:

$$\mathbf{x}_s = \underset{\mathbf{x}_i}{\operatorname{argmax}} d_{\mathbf{x}}(\mathbf{x}_i). \tag{8}$$

This methodology can be characterized as a passive sampling scheme, since the selection does not require values of the output variables, meaning that it shares some of the same advantages of the random scheme. As the points are added gradually, the distribution of points generates a hierarchical grid somewhat similar to the one generated by a grid search.

### 2.2.3. Greedy sampling on the output

In contrast to the previous strategy, greedy sampling on the output aims to find the unlabeled instance that maximizes the distance of estimated responses to the previously observed function values. The distance of an unlabeled instance in output space is given by:⋆

$$d_f(\mathbf{x}_i) = \min_j |f^\star(\mathbf{x}_i) - f(\mathbf{x}_j)|. \tag{9}$$

The next sample is selected according to:

$$\mathbf{x}_s = \underset{\mathbf{x}_i}{\operatorname{argmax}} d_f(\mathbf{x}_i). \tag{10}$$

This strategy will lead to a large number of samples in the regions where the response function varies rapidly, but potentially small exploration of the parameter space. In the case of multimodal functions, for example, this could lead to important features being missed in the sampling procedure.

### 2.2.4. Greedy sampling on both inputs and output

The greedy algorithm based on inputs and outputs is a combination of the previous criteria. The distances in both feature and response spaces are combined. The next sample is selected according to:

$$\mathbf{x}_s = \underset{\mathbf{x}_i}{\operatorname{argmax}} d_{\mathbf{x}}(\mathbf{x}_i) d_f(\mathbf{x}_i). \tag{11}$$

The distances in input and output spaces are calculated as shown before. The product of distances is used here instead of a weighted sum so that the resulting function is independent of the scales of the inputs and output.

### 2.2.5. Variational sampling

In the case of GP regressors, the standard deviation $\sigma^\star$ of the predictive distribution may be used as a criterion for choosing sampling locations. In this case, the sample is chosen with:

$$\mathbf{x}_s = \underset{\mathbf{x}_i}{\arg\max}\, \sigma^{\star}(\mathbf{x}_i). \qquad 12$$

The standard deviation does not depend directly on the response values obtained at the training stage, which is a characteristic of the Gaussian distribution (Rasmussen and Williams, 2005). However, these values are still utilized for the optimization of the kernel's hyperparameters, which affect in turn the expected covariance.

### 2.3. Computational fluid dynamics simulations

In order to quantify the performance of the systems analyzed here, the flow field inside the domain of interest has to be characterized. Engineering quantities such as pressure and shear stresses, for example, can be obtained from the knowledge of velocity and pressure fields, along with the fluid properties. The core equations to be solved are the continuity and momentum balance equations (Batchelor, 2000). The continuity equation for incompressible flows is given by:

$$\nabla \cdot \mathbf{u} = 0, \qquad 13$$

while the momentum equation is expressed by:

$$\frac{\partial \mathbf{u}}{\partial t} + \nabla \cdot (\mathbf{u} \otimes \mathbf{u}) - \nabla \cdot (v_{app} \nabla \mathbf{u}) = -\nabla p, \qquad 14$$

where $\mathbf{u}$ is the velocity, $p$ is the static pressure divided by the constant fluid density, $v_{app}$ is the apparent kinematic viscosity, and $t$ denotes time; gravitational forces have been neglected. We consider, for generality, that the apparent kinematic viscosity follows a power law $v_{app} = k|\dot{\gamma}|^{n-1}$, where $k$ is the consistency and $n$ is the flow index. The Newtonian behavior of the fluid, as assumed in test Cases 1 and 2, can be recovered by setting the value of $n$ to unity in which case we set $k \equiv v$, a constant. In nondimensional form, Equations (13) and (14) may be rewritten as:

$$\nabla \cdot \mathbf{u}^{\star} = 0 \qquad 15$$

and

$$\frac{\partial \mathbf{u}^{\star}}{\partial t} + \nabla \cdot (\mathbf{u}^{\star} \otimes \mathbf{u}^{\star}) - \frac{1}{Re} \nabla \cdot \left( \frac{v_{app}}{v_{ref}} \nabla \mathbf{u}^{\star} \right) = -\nabla p^{\star}, \qquad 16$$

where $Re$ ($= UL/v_{ref}$) is a Reynolds number in which $L$ and $U$ represent characteristic length and velocity scales, the choice of which is case dependent; here, $t$ and $p$ were scaled on $L/U$ and $\rho U^2$, respectively, wherein $\rho$ denotes the fluid (constant) density. The reference viscosity is taken to be $v_{ref} = k(U/L)^{n-1}$.

In the case of turbulent flows, two-equation models based on the turbulent viscosity are used such as the $k$-$\varepsilon$ model (Launder and Spalding, 1974). The simulations were performed with OpenFOAM-6 with two-dimensional setups, except for the last test case, which is fully three-dimensional (3D). In the case of axisymmetric flows (Cases 1 and 2), the geometry is defined as a wedge of small angle and thickness of a single cell in the out-of-plane direction. All simulations utilized second-order discretization schemes for the spatial terms, and steady state was assumed. If the mixing performance is to be evaluated (as in test Case 1), a transport equation for a passive scalar $c$ can also be solved:

$$\frac{\partial c}{\partial t} + \nabla \cdot (\mathbf{u}c) - \nabla \cdot \left( \frac{v_{eff}}{Sc} \nabla c \right) = 0, \qquad 17$$

where the turbulent viscosity, $v_{eff}$, is provided by the turbulence model and the Schmidt number, $Sc$, is taken to be unity.

## 3. Case Definitions

### 3.1. Case 1: flow through a static mixer

Industrial high efficiency vortex static mixers are commonly used for turbulent mixing along pipelines (Thakur et al., 2003; Eissa, 2009). The first case in this study aims to model the geometry of such a mixer.

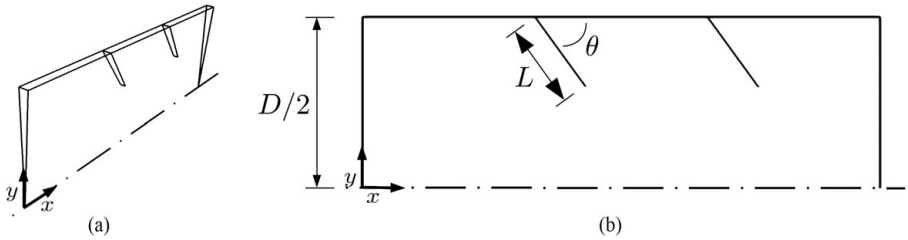

***Figure 2.*** *Schematic representation of the static mixer geometry used in Case 1 with side- and front-views shown in (a) and (b), respectively.*

The fluid enters the domain through the leftmost boundary, where a passive scalar is inserted in part of the domain. As it goes across the device, turbulent mixing takes place, eventually leading to a nearly uniform concentration of the tracer on the cross-section. To evaluate the mixing performance of the device, Equation (17) was solved along with the flow equations, as mentioned previously. The two-dimensional geometry of the static mixer is shown in Figure 2. For the surrogate model, we mainly focus on two of the parameters: blade length to diameter ratio ($L/D$) from 0.2 to 0.3, and blade angle ($\theta$) from 0.8 to 1.57 rad. The Reynolds number, $Re$ ($= UD/v$), is kept constant at $10^5$, where $U$ is the constant speed of the fluid that enters the mixer from the left boundary, where a uniform velocity profile is imposed; note that here we consider the fluid to be Newtonian with a constant kinematic viscosity, $v$. For all the walls in contact with the fluid (e.g., upper wall of the wedge, and blade walls), a no-slip condition is imposed. For scalar transport, the tracer injection at the inlet is set as a step profile similar to the one presented by (Eissa, 2009). The turbulence model used is the standard $k$-$\varepsilon$ (Launder and Spalding, 1974), and a turbulence intensity of 5% was set for the inlet. To determine the mixing performance, $c_v$ at the outlet is calculated, which corresponds to the variance of the tracer concentration at the outlet:

$$c_v = \frac{\sqrt{\overline{(c - \overline{c})^2}}}{\overline{c}},\tag{18}$$

where the average concentration, $\overline{c}$, is computed over the outlet area.

In order to assure that the results were valid in a wide range of conditions, the wall boundary conditions utilized for $\varepsilon$, $k$, and $v_t$ were set to, respectively *epsilonWallFunction*, *kLowReWallFunction*, and *nutUSpaldingWallFunction*, which automatically switch between low- and high-Reynolds number formulations as needed.

Figure 3 shows the nondimensional velocity magnitude, turbulent kinetic energy, and concentration fields obtained for a geometry similar to the one presented by Eissa (2009), with an angle of $\theta = \pi/4$. Despite the fact that the current results were generated using a two-dimensional formulation, they are nonetheless in qualitative agreement with the results of Eissa (2009) and Bakker and Laroche (2000). One major discrepancy with the work by Eissa (2009) is the cross-sectional vortices generated at the blade tips. In order to reproduce those more accurately, it is possible for us to deploy more sophisticated turbulence models, or large eddy simulations, to better capture these phenomena, but this is not the focus of the present work.

### 3.2. Case 2: orifice flow

Flow through an orifice is a common occurrence in industry with applications that include pipeline transportation of fluids, and flow in chemical reactors and mixers wherein orifice flows are used for the purpose of flow rate measurement and regulation. Here, pressure drop across the orifice as a function of system parameters is investigated. As the shape is symmetrical in both the radial and axial directions, the computational domain is a pipe sector, as shown in Figure 4. A uniform velocity profile is considered for the inlet boundary, and no slip conditions are applied for the wall. At the outlet, a constant pressure is assumed. When creating the surrogate model, the geometry of the orifice is allowed to vary, such that each configuration is defined by the area ratio $(d/D)^2$ and length to diameter ratio $s/d$. The Reynolds number $Re$, given here by $UD/v$, is taken to be constant and equal to $10^5$. The flow field simulated in OpenFOAM is using the SST $k$-$\omega$ model (Menter et al., 2003). A turbulence intensity of 2% was assumed for calculation of turbulent kinetic energy at the inlet, and this value was used along with a prescribed mixing length for definition of the specific turbulence dissipation value. The chosen mesh resolution resulted in a total of $5.1 \times 10^4$ cells for a setup with area ratio $(d/D)^2 = 0.54$ and length to diameter ratio $s/d = 0.2$ (see Figure 11). Mesh refinement is utilized to increase the mesh density close to wall while reducing computational cost simultaneously. A mesh-refinement study has been conducted to prove that the solutions obtained are mesh-independent. The difference in pressure drop across the *vena contracta* between $2.5 \times 10^4$ cells and $5.1 \times 10^4$ cells is less than 2% as well as the difference between $5.1 \times 10^4$ cells and $10^5$; thus the setup for $5.1 \times 10^4$ cells was deemed to produce simulation results that are unaffected by mesh resolution.

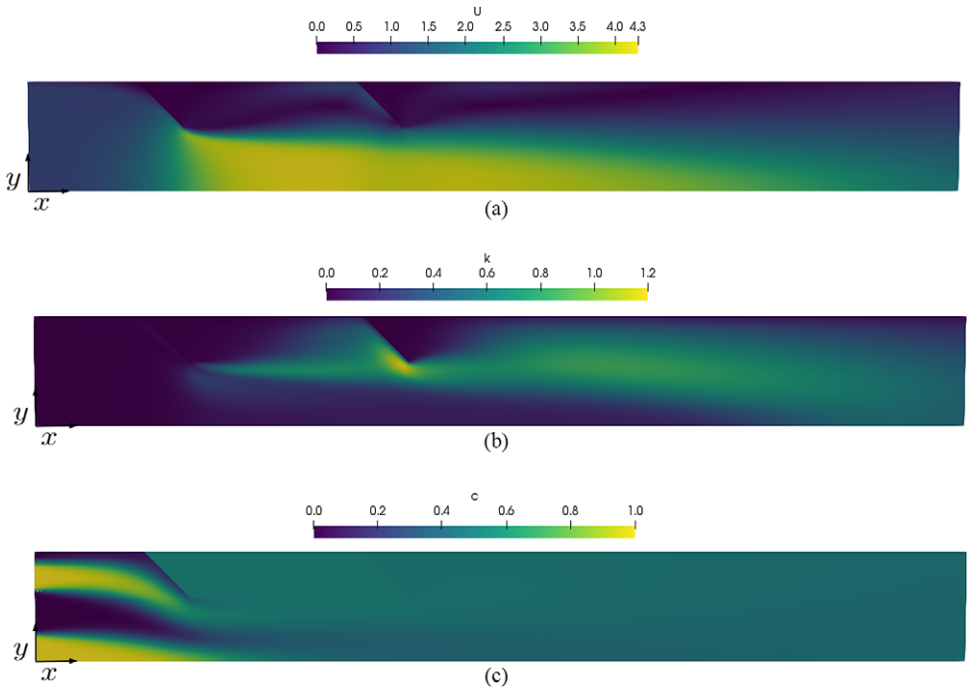

**Figure 3.** *Computational fluid dynamics simulation of turbulent scalar transport through a static mixer using the k-ε model (Launder and Spalding, 1974) for Case 1 showing steady, two-dimensional nondimensional velocity magnitude (a), turbulent kinetic energy (b), and scalar concentration (c), fields, generated with $Re = 10^5$, $L/D = 0.3$, and $\theta = \pi/4$. The scale bars represent the magnitude of the fields depicted in each panel.*

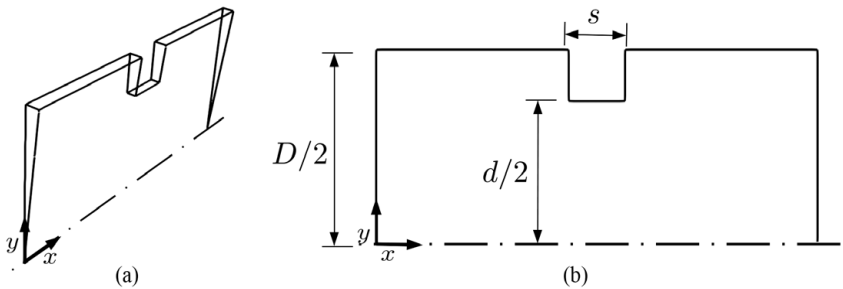

**Figure 4.** *Schematic representation of the orifice geometry used in Case 2 with side- and front-views shown in (a) and (b), respectively.*

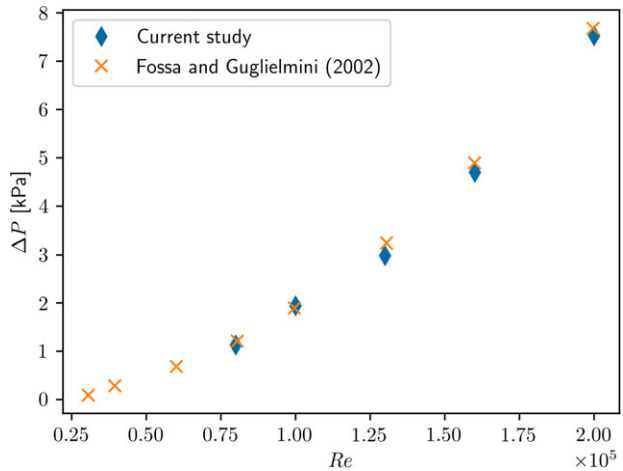

**Figure 5.** *A comparison of the current predictions of pressure drop across the orifice as a function of the local Reynolds number with the experimental data of Fossa and Guglielmini (2002) with $(d/D)^2 = 0.54$ and $s/d = 0.20$ (see Figure 4b).*

Since the change in geometry is expected to have an impact on the mesh requirements for the wall, the boundary conditions for $\omega$, $k$, and $\nu_t$ were set to, respectively *omegaWallFunction*, *kLowReWallFunction*, and *nutUSpaldingWallFunction*, which are able to deal with low- and high-Reynolds number scenarios as needed.

In Figure 5, we compare our results with the of experimental data of Fossa and Guglielmini (2002) for the local pressure drop across the orifice as a function of the flow Reynolds number. Inspection of Figure 5 reveals excellent agreement—with maximum deviations of 7.9%—which inspires confidence in the reliability of the present numerical predictions.

### 3.3. Case 3: flow in an inline mixer (2D)

The third case is representative of flow of Newtonian and power-law fluids in an inline mixer reported by Vial et al. (2015). This setup utilizes a two-dimensional geometry, neglecting the effect of the flow in the axial direction. The geometry and key variables are shown in Figure 6. The impeller rotates in the counter-clockwise direction around the *z*-axis, while the outer tank wall is kept fixed. For further efficiency gains in the simulation, a periodic symmetry is prescribed in the tangential direction. For this system, the power number (per nondimensional length $L_e/d$) $N_{p,2D}$ is chosen as the main variable of interest. It is given by:

$$N_{p,2D} = \frac{2\pi T_{2D}}{N^2 d^4}, \tag{19}$$

where $T_{2D}$ is the torque applied per fluid mass and $N$ is the rotation speed. Through dimensional analysis, it can be shown that this value is a function of the Reynolds number, $Re$ ($= d2\pi N^{(2-n)}d^2/k$), nondimensional gap between rotor and stator $\alpha$ ($= (D-d)/D$), flow index $n$, and number of blades $N_b$.

The simulations performed were two-dimensional. Vial et al. (2015) reported that this simplification was adequate to describe the behavior of the system as long as the flow was in the laminar regime and the nondimensional gap between rotor and stator was kept below 0.2. Thus, these restrictions were considered in the choice of parameter ranges. Similarly to the original study, the computational domain consisted of a single blade, and cyclic boundaries were applied for the tangential direction. No-slip conditions were considered for the walls, and the rotation effect was included through the Single reference frame (SRF) model. Further details about the SRF methodology can be found in Wilhelm (2015).

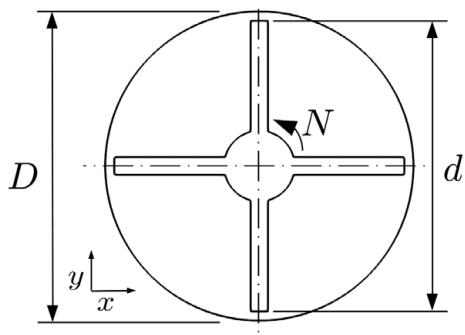

**Figure 6.** *Schematic representation of the geometry used in Case 3, two-dimensional flow in an inline mixer.*

All computations were performed with a quadrilateral mesh with circumferential resolution $\Delta s/D_s = 0.005$ and radial resolutions $\Delta r_i/D_s = 0.005$ and $\Delta r_o/D_s = 0.002$ at the rotor and stator regions, respectively. The difference in the power number obtained with a finer mesh, with double the resolution, was 0.3%. Table 2 shows the comparison of the power coefficient $k_p$ $(= N_{p,2D} Re L_e/d)$ calculated here and the results obtained by Vial et al. (2015) through experiments and two-dimensional simulations. The agreement is very good.

### 3.4. Case 4: flow in an inline mixer (3D)

The final test case is a 3D version of the inline mixer presented before. Here, besides the rotation around the *z*-axis, we also take into account the flow along that direction, as well as the finite length of the blades. The newly introduced variables are illustrated along with the general geometry in Figure 7.

Besides the parameters defined before, this setup introduces two new parameters to the input: the axial Reynolds number $Re_{ax}$ $(= dUN^{(1-n)}d/k)$, and the nondimensional length $L_e/D$, for a total of six parameters. The mesh utilized for the 3D case was less refined than for the two-dimensional setup, with circumferential resolution $\Delta s/D_s = 0.01$, radial resolutions $\Delta r_i/D_s = 0.01$ and $\Delta r_o/D_s = 0.005$ at the rotor and stator regions, respectively, and axial resolution $\Delta z/D_s = 0.01$. The power number difference to a mesh with double the resolution was 1.2%. The power coefficient numbers obtained with this setup and the comparison to the experimental values are shown in Table 2. The values are well within the experimental uncertainty of the measurements.

### 3.5. Summary of regression variables

Table 3 shows a summary of the parameters considered for the surrogate models, along with the respective output variables. Note that all variables are nondimensional, in order to enforce similarity constraints. The variables that could vary over several orders of magnitude were represented in logarithmic form.

**Table 2.** Comparison between present results for power coefficient, $\alpha$, and those reported by Vial et al. (2015).

| $\alpha$ | Exp. (Vial et al., 2015) | Vial et al. (2015)—2D | Present—2D | Present—3D |
|---|---|---|---|---|
| 0.057 | $3100 \pm 400$ | 3,000 | 2,950 | 3,030 |
| 0.2 | $2000 \pm 300$ | 1,680 | 1,690 | 1,800 |

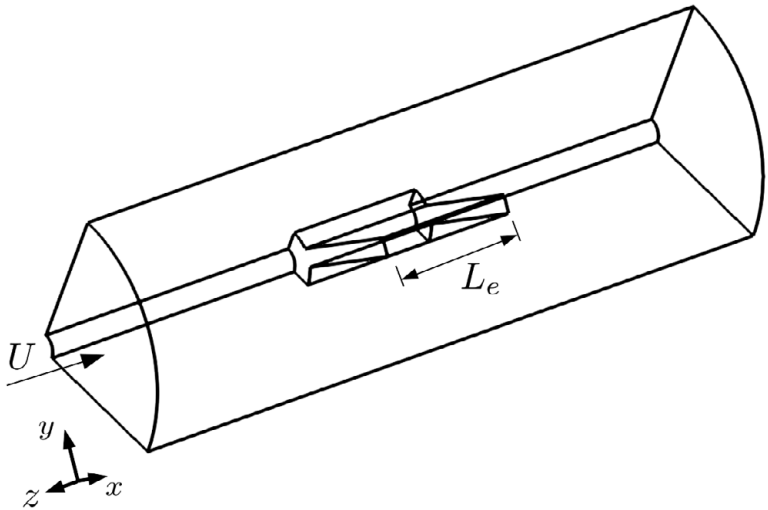

**Figure 7.** *Schematic representation of the geometry used in Case 4, three-dimensional flow in an inline mixer.*

**Table 3.** Features and responses utilized in each of the three test cases.

| Case | 1 | 2 | 3 | 4 |
|---|---|---|---|---|
| $x_1$ | $L/D$ | $(d/D)^2$ | $\log Re$ | $\log Re$ |
| $x_2$ | $\theta$ | $s/D$ | $\alpha$ | $\alpha$ |
| $x_3$ | – | – | $n$ | $n$ |
| $x_4$ | – | – | $N_b$ | $N_b$ |
| $x_5$ | – | – | – | $Re_{ax}$ |
| $x_6$ | – | – | – | $L_e/D$ |
| $f(x)$ | $\log c_v$ | $\frac{\Delta P}{\rho U^2}$ | $\log N_{p,2D}$ | $\log N_{p,2D}$ |

Table 4 presents the comparison between the average computational costs of the test cases, including the time to both generate the parameterized mesh as well as to calculate the solution of the flow. The values range from a few minutes to 1 h, and are directly related to the size of the grid that is required.

## 4. Results and Discussion

### 4.1. Performance evaluation methodology

Before the active-learning techniques are employed, an initial version of the regression model must be calibrated. All methods are initialized with the procedure proposed by Wu et al. (2019): the first point is located at the centroid of the parameter space, followed by $N_0 - 1$ locations selected with the passive strategy of greedy sampling on inputs. We set the total number of initial simulations $N_0$ to be equal to $4N_{\text{features}}$ in Cases 1 and 2, $2N_{\text{features}}$ in Cases 3 and 6 in Case 4.

In order to quantify the global interpolation performance of each algorithm, a set of $N_e$ simulations with parameters $\mathbf{x}_i$ is performed, independent of the ones used in the development of the surrogate model, to be used as "ground truth." The average relative error is given by Equation (20):

**Table 4.** Typical computational cost of individual simulations of each test case.

| Case | Number of cores | Time [s] |
|---|---|---|
| 1 | 1 | 260 |
| 2 | 1 | 1,070 |
| 3 | 1 | 1,020 |
| 4 | 192 | 3,430 |

$$\varepsilon^{\star} = \frac{1}{N_e} \sum_{i=0}^{N_e} \frac{|f(\mathbf{x}_i) - f^{\star}(\mathbf{x}_i)|}{|f(\mathbf{x}_i)|}. \tag{20}$$

The positions $\mathbf{x}_i$ for error estimations were chosen randomly, with the total number of simulations $N_e$ set to 100. Additionally, for the techniques that are stochastic—either due to the regression or sampling methods—10 repetitions were performed. Results are presented in the form of a mean value and a confidence interval of 95% based on the *t*-distribution. Due to the very high computational cost of the full 3D computations, we have utilized a single repetition of each for Case 4, such that no confidence bars are provided.

### 4.2. Discussion

Figure 8 presents a qualitative comparison of the interpolation obtained for Case 1 using the regression techniques presented in Section 2.1. In order to better isolate effects, a fixed sampling method was used for these results: in this case, greedy I/O, based on the recommendation of Wu et al. (2019). The predictions of all methods are found to be qualitatively very similar. The notable outlier is the random forest regression —although the predicted values are reasonable, the surface is non-smooth, since the values originate from a combination of decision trees. It is important to highlight that, despite using the same sampling strategy, the sampling locations for each regression method may be different, since they also depend on the predicted values provided by the regression.

In Figure 9, a qualitative visual comparison between sampling strategies for case 1 is presented. Based on preliminary performance tests which indicated it performed well in quantitative tests, the fixed regression strategy chosen here was a Gaussian process with the Matérn 5/2 kernel. A good sampling strategy should be able to adapt to both the situation of "effect sparsity"—where only some of the parameters significantly affect the solution—as well as the situation where all inputs have equal importance (Morris and Mitchell, 1995). As such, sampling positions should be spread out enough so as to fully capture the main features of the response surface, while simultaneously avoiding unnecessary computations in regions of low gradient. The greedy input and variational methods provide a uniform coverage of the domain, while the greedy output sampling is heavily biased towards the center and top right regions. The greedy I/O, as a mix of two criterion, has a somewhat attenuated bias.

Figure 10 presents the calculated interpolation error for each regression strategy. The relative errors are high for Case 2 since the response function has values close to zero in a large part of the domain, but the trends are somewhat consistent between all cases. Since it has very few degrees of freedom, the strategy employing linear regression rapidly stagnates after a small number of simulations. Similarly, the performance of the multilayer perceptron does not improve significantly as samples are added; most likely, this can be attributed to overfitting due to the quite limited number of data points. The Gaussian process regression performs well in all three cases, while Support Vector regression has mixed results: the performance is satisfactory for Cases 1 and 3 but poor for Case 3. Finally, it is noticeable that, in general, the curves are nonsmooth and nonmonotonic, meaning that there is room for significant optimization with the choice of sampling quantity and locations.

Figure 11 show the progress of the estimated error as the number of simulations increases, for different sampling strategies. For Cases 1 and 2, most of the examined strategies had performance comparable or slightly better than the random baseline. For the latter, in particular, the errors obtained with the variational, greedy I and greedy I/O strategies were lower than half of the baseline values, with more than approximately five samples. For Case 3, the variational and greedy I significantly outperformed random sampling in the range of 5–30 samples. Despite the relatively low dimensionality of the the cases that were examined, some positive effect may be observed already with the use of the passive or active strategies, especially at low numbers of samples.

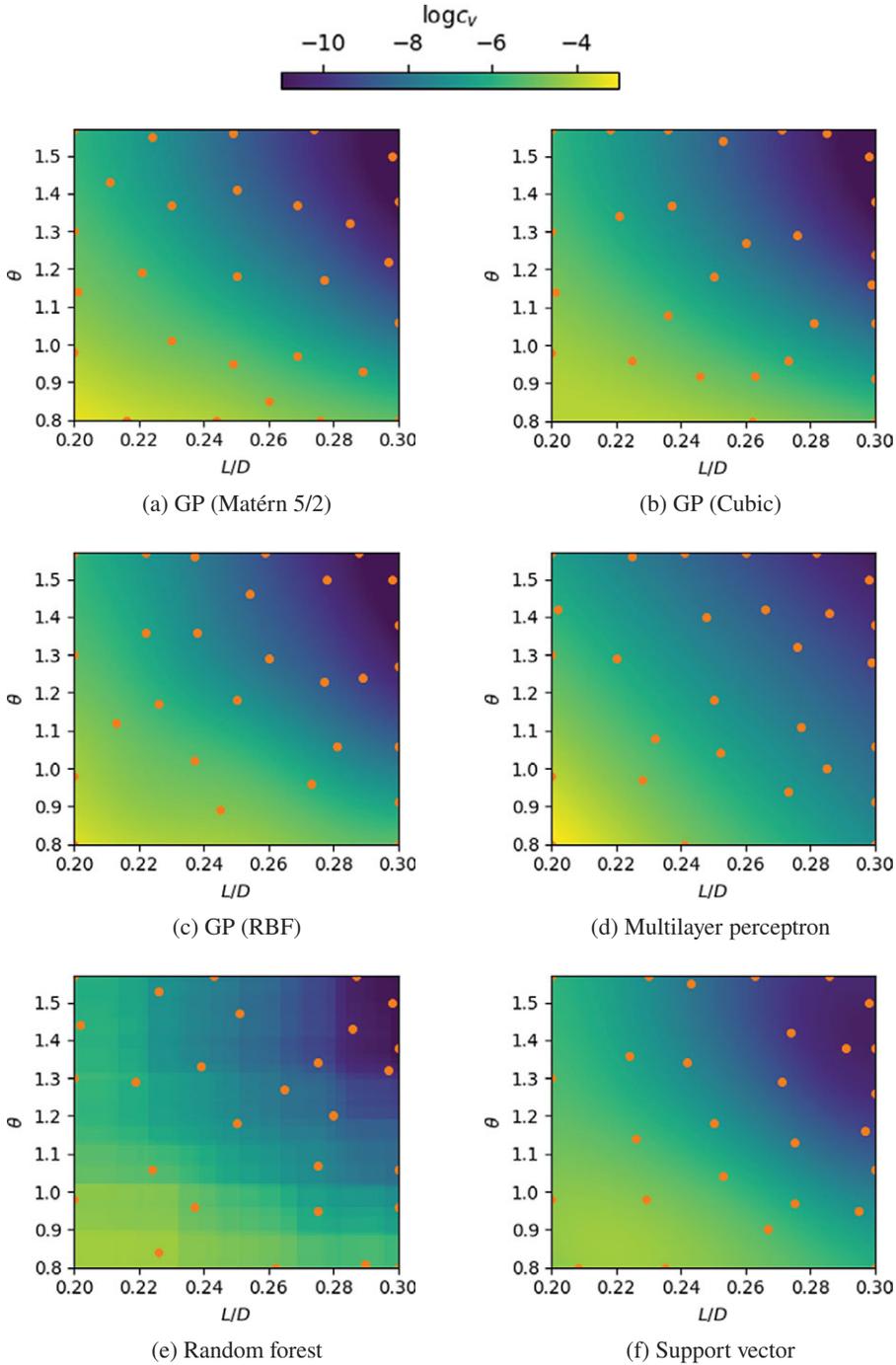

**Figure 8.** *Comparison between regressions for case 1, with 20 samples. Circles indicate sampling locations.*

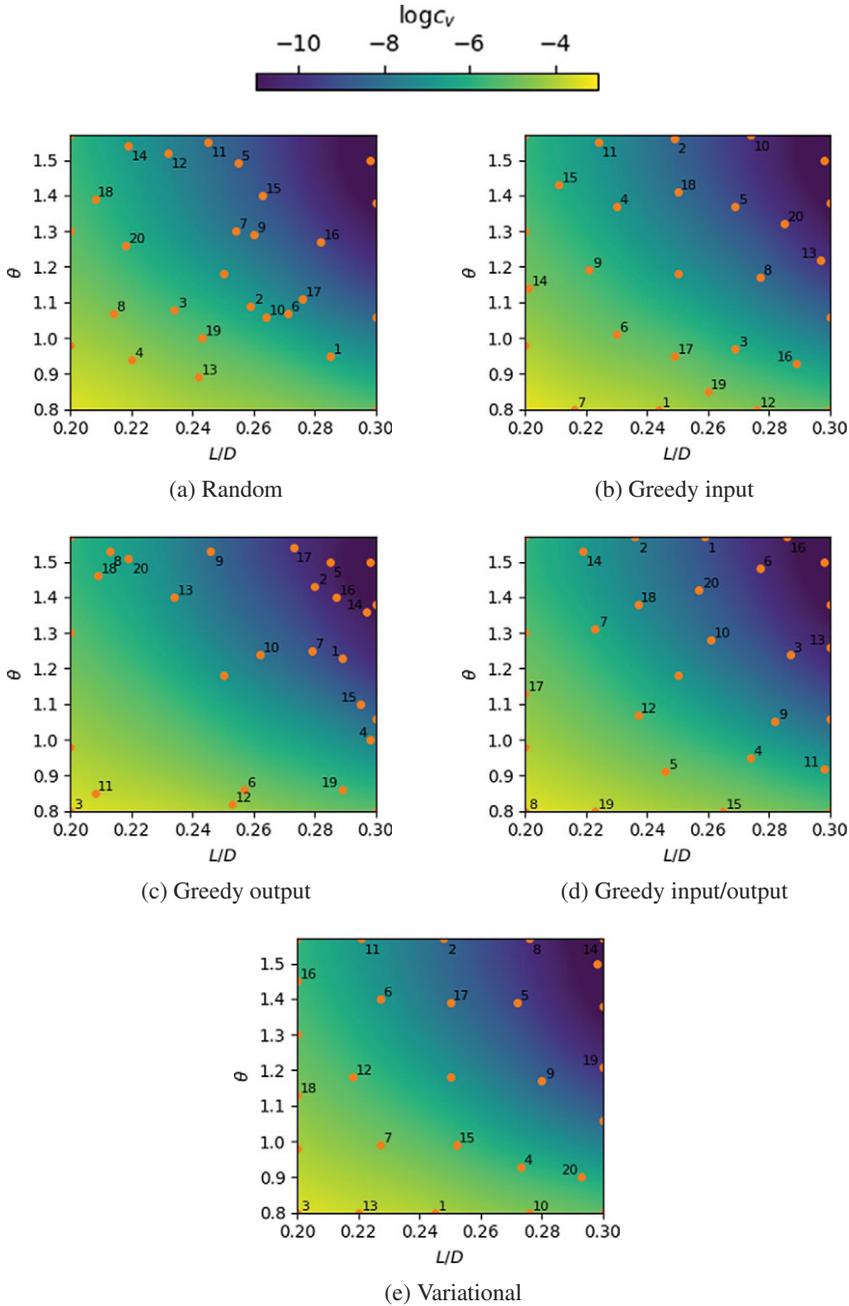

**Figure 9.** *Comparison between sampling strategies for Case 1, with 20 samples. Circles indicate sampling locations and numbers indicate the sampling order (positions used for initialization are unlabeled).*

The scalability of the techniques to more complex systems can be evaluated by analyzing Figure 12, which presents the performance of the regression and sampling techniques as a function of the number of simulations performed on Case 4. The general trend is somewhat similar to the one observed for Case 3, a similar mixer setup. However, due to the limited samples and larger parameter space, instances of severe overfitting are observed in some situations, leading to predictions outside of the training data to be completely erroneous. In fact, the only sampling technique that seems to be able to provide reasonable predictions for the GP model is the GIO algorithm.

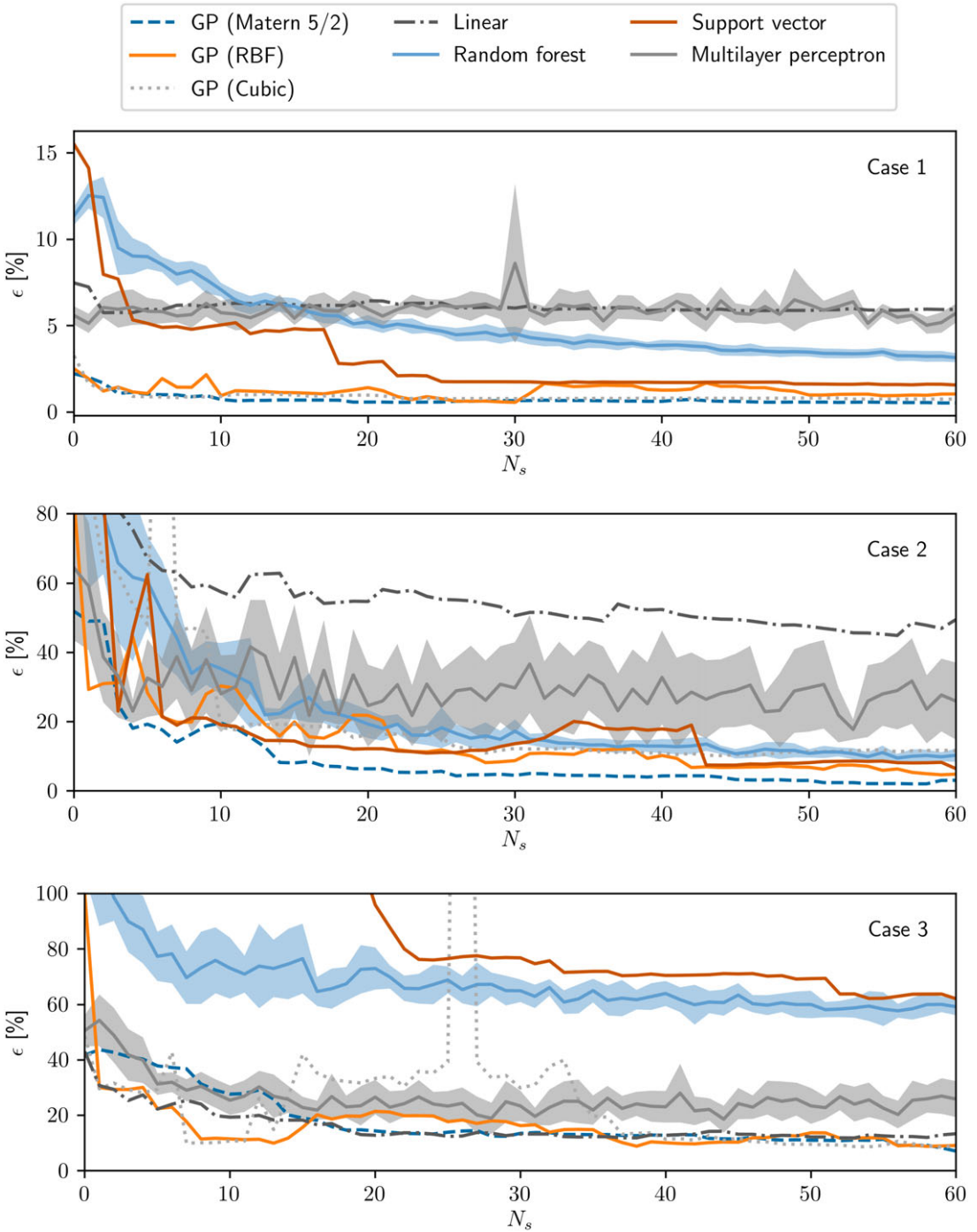

**Figure 10.** *Error as a function of number of samples beyond the ones used for the initialization for different regression strategies, for case 1 (top), case 2 (middle) and case 3 (bottom).*

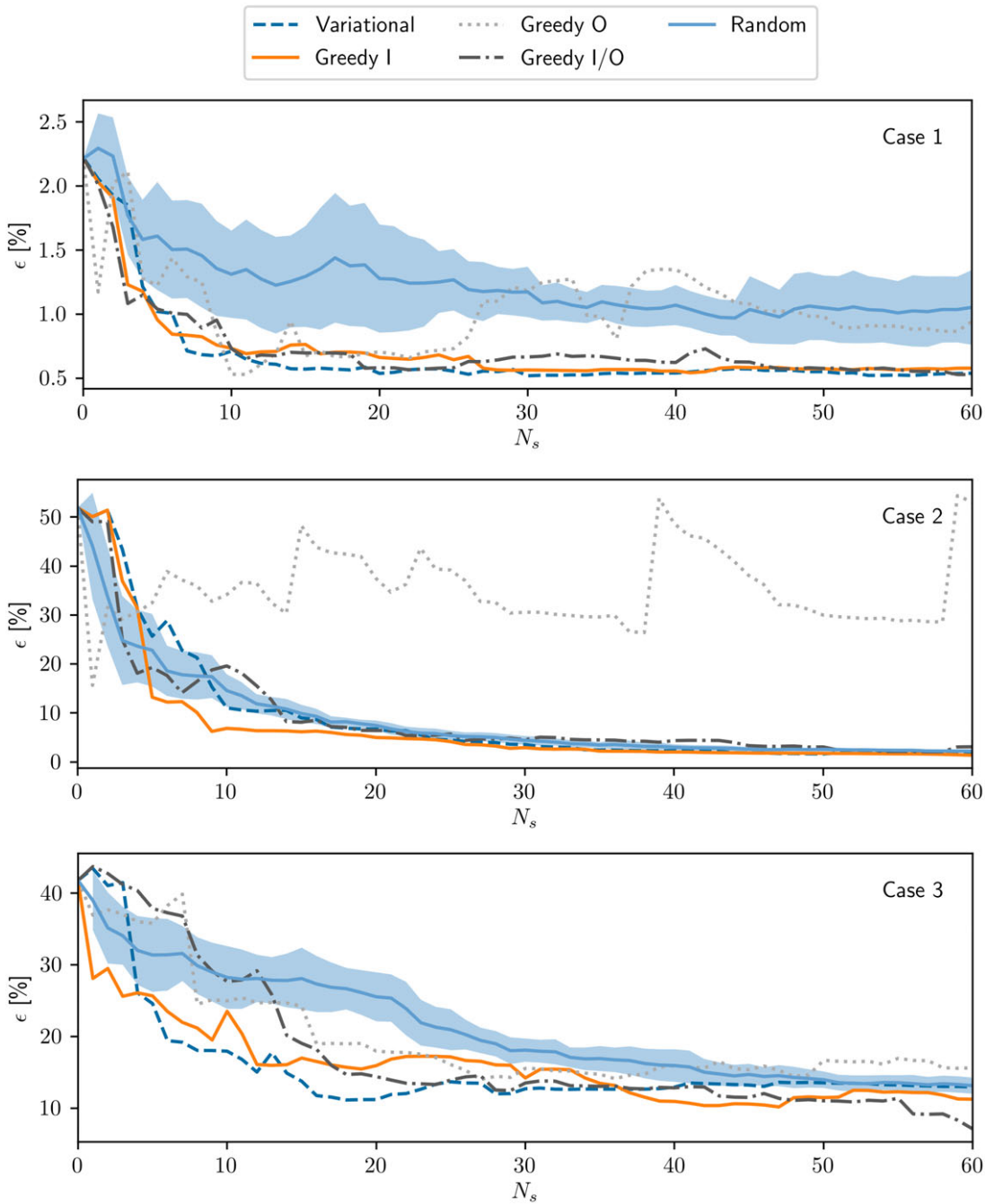

**Figure 11.** *Error as a function of number of samples beyond the ones used for the initialization for different sampling strategies, for Case 1 (top), Case 2 (middle), and Case 3 (bottom).*

The overfitting problem is illustrated explicitly in Figure 13, which compares the predictions for training and test data, for two different sampling methodologies. The GP fit is quite flexible and able to reproduce the set of training data very well. In complex cases, however, the hyperparameter optimization may lead to length scales which are too small, such that the regression function collapses into a

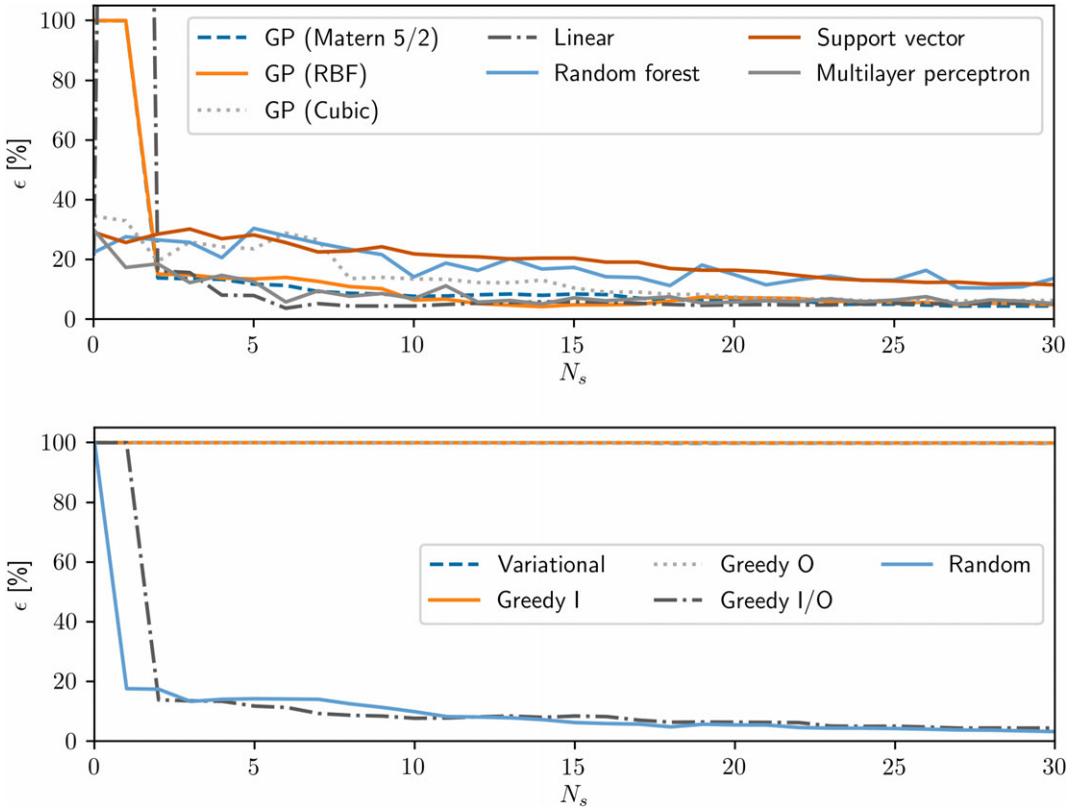

**Figure 12.** *Error of the regressions developed for Case 4 as a function of number of samples beyond the ones used for the initialization, for different regression strategies (top), and different sampling strategies (bottom).*

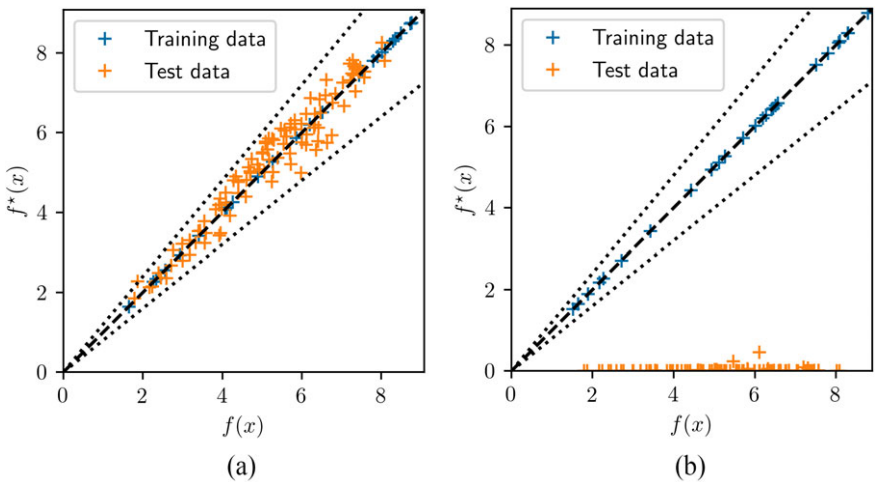

**Figure 13.** *Comparison between predicted and reference values for the GP52 regression with GIO (a) and variational sampling (b) generated with 20 samples for Case 4.*

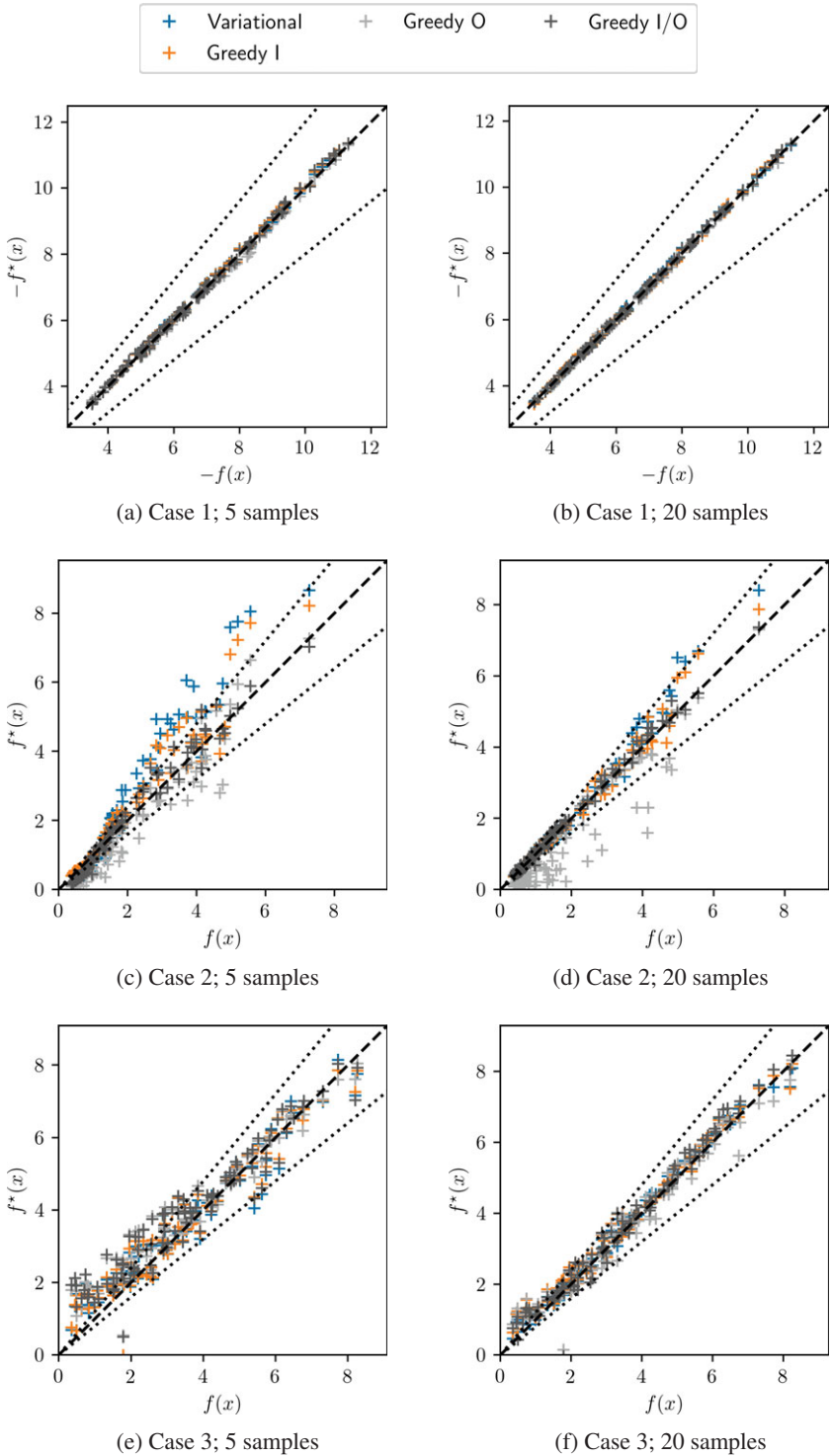

**Figure 14.** *Comparison between predicted and reference values for each Cases 1–3, with 5 (left) or 20 samples (right). Dashed lines represent a 20% error.*

**Table 5.** Error for different sampling strategies, for 5, 10 or 20 queries.

| Case | Queries | Strategy | | | | |
|---|---|---|---|---|---|---|
| | | V | GI | GO | GIO | R |
| 1 | 5 | 1 | **0.96** | 1.2 | 1 | 1.6 [1.2–2.0] |
| | 10 | **0.71** | 0.73 | 0.53 | 0.73 | 1.3 [0.97–1.7] |
| | 20 | **0.54** | 0.66 | 0.7 | 0.58 | 1.3 [0.85–1.7] |
| 2 | 5 | 26 | **13** | 32 | 19 | 23 [15–30] |
| | 10 | 11 | **6.8** | 34 | 20 | 15 [11–18] |
| | 20 | 6.7 | **5** | 37 | 6.4 | 7.5 [6.5–8.5] |
| 3 | 5 | **25** | 26 | 36 | 38 | 31 [26–37] |
| | 10 | **18** | 24 | 25 | 28 | 28 [24–33] |
| | 20 | **11** | 16 | 18 | 14 | 26 [22–29] |
| 4 | 5 | 100 | 100 | 100 | **12** | 14 |
| | 10 | 100 | 100 | 100 | **7.7** | 9.9 |
| | 20 | 100 | 100 | 100 | 6.4 | **5.5** |

Bold values represented the best performance in each configuration.
Abbreviations: GI, greedy sampling on the inputs; GIO, greedy sampling on both inputs and output; GO, greedy sampling on the output; R, random; V, variational.

combination of singularities around the training data. This results in meaningless values for any new data outside the immediate neighborhood of the training set. Although this could be addressed through fine tuning of the algorithm, it would require a case-by-case analysis. In this case, the GIO strategy seems to achieve a balance between density and coverage of the sampling space, leading to a smoother convergence of the length scale to an adequate value.

Similar comparisons of predictions and reference values are shown for the other cases on Figure 14. Systematic deviations may be observed in some situations for variational, GI, and GO sampling. Using a more balanced sampling approach such as GIO, which does not focus on one specific area of the parameter space seems to aid the most stable result outcome without the need of additional fitting procedures.

A summary of the error values obtained with each sampling strategy, for selected numbers of samples (5, 10, or 20), is shown in Table 5. The active learning strategies managed to outperform random sampling in most of the chosen configurations. The variational method, based on estimates of standard deviation provided by the GP, presented the best performance on Cases 1 and 3, followed closely by the greedy input passive sampling. On Case 2, the latter performed particularly well, although the variational approach also presented satisfactory results in the case of 10 and 20 samples. Greedy sampling on the outputs strategy proved inconsistent across the three two-dimensional cases, where it performed particularly bad for Case 2. Once more the last remaining greedy strategy (GIO) proved to work persistently across the first three cases, whereas for Case 4 it was the only algorithm that provided satisfactory results. In the 3D case GIO outperformed random sampling for low query numbers, whereas the latter began to be increasingly competitive as the number of samples was increased.

## 5. Final Remarks and Future Directions

We have evaluated different active-learning strategies for the development of surrogate models based on CFD simulations. The procedure was applied to a selection of industrially-relevant flow cases, and an extensive comparison between different regression and sampling algorithms was performed. In general, one active learning strategy (GIO) proved to work consistently across all cases, outperforming random sampling in most instances. This strategy proved superior to other active learning strategies particularly for Case 4, where it provided meaningful results without the need for additional fine tuning of the fitting algorithm. For the lower dimensional (Cases 1–3), two active-learning strategies (variational and greedy sampling on inputs) outperformed random sampling for all cases considered.

The very small sample sizes, in this study, due to the restriction on the number of queries ensures that only a limited part of the parameter space for low-dimensional cases is explored, which is also the case in high-dimensional problems even with increased samples. Hence potentially similar results could hold for higher-dimensional problems, but this needs to be quantified in future studies. Using the methodology presented here, further benchmarks need to be developed for other classes of flow problems, involving, for example, discontinuous parameter spaces and/or noisy or stochastic results. The algorithms could also take into account the non-uniform computational cost of the simulations using an approach similar to that followed by Snoek et al. (2012). To improve upon the present results in future work, the CFD-agnostic active-learning schemes employed in the present paper might need to incorporate physical insights and case-specific knowledge.

**Funding Statement.** We acknowledge Funding from the UK Research and Innovation, and Engineering and Physical Sciences Research Council through the PREdictive Modeling with QuantIfication of UncERtainty for MultiphasE Systems (PREMIERE) program, grant number EP/T000414/1, the AI for Science and Government project on Digital Twins for Multiphase Flow Systems, and the Royal Academy of Engineering for a Research Chair in Multiphase Fluid Dynamics for OKM. G.G. acknowledges funding by CNPq, Brazil. I.P. acknowledges funding from the Imperial College Research Fellowship scheme.

**Competing Interests.** The authors declare no competing interests exist.

**Data Availability Statement.** Scripts and test cases setups may be found in the following repository: https://github.com/ImperialCollegeLondon/al_cfd_benchmark.

**Ethical Standards.** The research meets all ethical guidelines, including adherence to the legal requirements of the study country.

**Author Contributions.** Conceptualization, L.M. and I.P.; Investigation, G.G., Y.L., and Y.L.; Software, G.G.; Supervision, A.B., I.P., and O.M. Writing-original draft, G.G., Y.L., and Y.L.; Writing-review & editing, A.B., L.M., I.P., and O.M. All authors approved the final submitted draft.

## Appendix

### Mesh Study of Cases 1 and 2

For Case 1, the effect of mesh resolution on the concentration field along the symmetry axis is presented in Figure A1. The chosen resolution proves sufficient for capturing the solution accurately without requiring excessive time for computation. For the base case, the mesh adopted had an average $y^+$ of 38.7 and 22.3 and maximum of 80.8 and 67.4 for walls and blades, respectively, which is adequate for the high-Reynolds model.

For Case 2, the pressure profiles along the symmetry axis for different mesh resolutions are presented in Figure A2. The results are nearly indistinguishable between the three meshes.

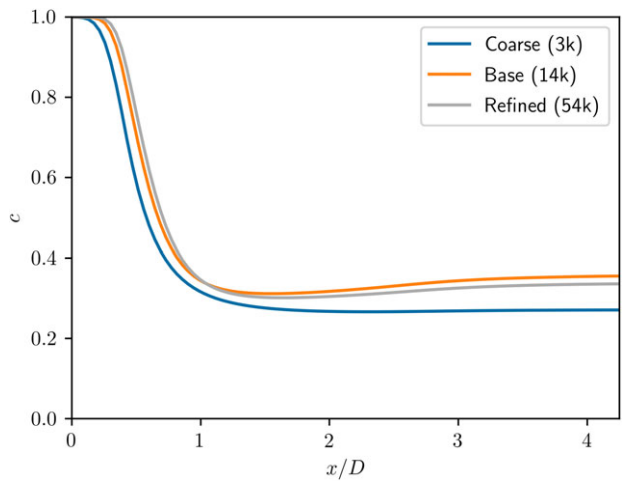

**Figure A1.** *Effect of mesh resolution on concentration profile along the symmetry axis.*

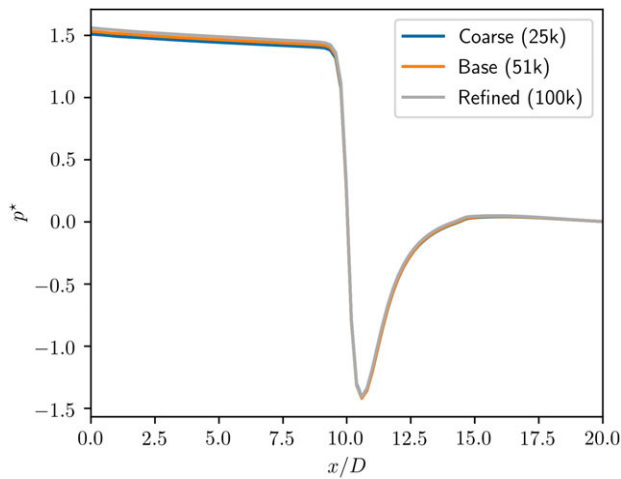

**Figure A2.** *Effect of mesh resolution on pressure profile along the symmetry axis.*